# Direct Simplified Symbolic Analysis (DSSA) Tool

Mohammad Shokouhifar[1,2,*], Hossein Yazdanjouei[3,4], Gerhard-Wilhelm Weber[5]

[1] [a] *Department of Electrical and Computer Engineering, Shahid Beheshti University, Tehran 1983969411, Iran*
[2] *Institute of Research and Development, Duy Tan University, Da Nang 550000, Vietnam*
[3] *Microelectronics Research Laboratory, Urmia University, Urmia, Iran*
[4] *Department of Computer Science, Khazar University, Baku AZ1096, Azerbaijan*
[5] *Faculty of Engineering Management, Poznan University of Technology, Poland*

**A B S T R A C T**

___________________________________________________________________________________

This paper introduces Direct Simplified Symbolic Analysis (DSSA), a new method for simplifying analog circuits. Unlike traditional matrix- or graph-based techniques that are often slow and memory-intensive, DSSA treats the task as a modeling problem and directly extracts the most significant transfer function terms. By combining Monte Carlo simulation with a genetic algorithm, it minimizes error between simplified symbolic and exact numeric expressions. Tests on five circuits in MATLAB show strong performance, with only 0.64 dB average and 1.36 dB maximum variation in dc-gain, along with a 6.8% average pole/zero error. These results highlight DSSA as an efficient and accurate tool for symbolic circuit analysis.

___________________________________________________________________________________

___________________________________________________________________________________

**\* Correspondence:** Mohammad Shokouhifar (ORCID: 0000-0001-7370-4760).

**Email addresses:**

Mohammad Shokouhifar: mohammadshokouhifar@duytan.edu.vn

Hossein Yazdanjouei: h.yazdanjouei@gmail.com

Gerhard-Wilhelm Weber: gerhard.weber@put.poznan.pl

## 1. Introduction

Symbolic analysis investigates analog circuits by representing circuit equations with symbolic variables, enabling a deeper understanding of the relationships between components beyond pure numerical methods. This approach is particularly valuable for complex or nonlinear circuits, allowing designers to algebraically manipulate equations and explore how changing component values affects circuit behavior [1]. However, performing precise symbolic analysis manually is often tedious and error-prone, even for small circuits. To address this, automatic symbolic analysis tools, such as MATLAB, GNU Octave, or MAPLE, use mathematical solvers like Cramer's rule and generally rely on graph- or matrix-based techniques [2].

Despite their utility, symbolic analysis methods often produce highly complex equations that are difficult to interpret, making simplification essential [3]. Simplification techniques are typically categorized by when they are applied: simplification-before-generation (SBG) on circuit models or matrices, simplification-during-generation (SDG) during analysis, and simplification-after-generation (SAG) on final symbolic expressions [4,5]. For real-world circuits, where transfer functions contain numerous terms, these simplification steps are crucial to extract meaningful insights [6]. Over the years, heuristic and metaheuristic methods have been proposed for symbolic simplification, yet they mostly operate on fully expanded expressions, which can generate substantial computational overhead for large circuits.

To overcome these limitations, we propose Direct Simplified Symbolic Analysis (DSSA), a metaheuristic-driven modeling approach that focuses on extracting only the most significant terms of the transfer function. Unlike conventional graph- or matrix-based analyses and SAG, SDG, or SBG simplifications, DSSA formulates the problem as an optimization task, ensuring numerical equivalence while avoiding unnecessary computational complexity. By utilizing a genetic algorithm (GA) and Monte Carlo-generated datasets that capture parameter variations and uncertainties, DSSA efficiently identifies essential terms while dramatically reducing execution time, memory use, and overall computational load.

The proposed DSSA tool advances symbolic analysis by: directly generating simplified equations from the circuit netlist; redefining symbolic simplification as an optimization problem; employing Monte Carlo data generation and GA-based optimization; and providing a MATLAB implementation tested on five circuits of varying complexity. This method significantly improves the efficiency, scalability, and practical applicability of symbolic analysis in transistor-level circuits.

The remainder of the paper reviews existing techniques (Section 2), details the DSSA method (Section 3), presents simulation results (Section 4), and concludes with final insights (Section 5).

## 2. Literature Review

Symbolic simplification plays a vital role in analog circuit analysis, offering significant improvements in efficiency and accuracy [7]. However, symbolic analysis is inherently NP-hard [1], making it computationally challenging for practical circuits. Calculating symbolic transfer functions using Kirchhoff laws at nodes and meshes can be time-consuming and complex [8]. To support designers, CAD tools such as MATLAB, GNU Octave, or MAPLE are often employed to extract symbolic functions. While manual circuit-level calculations are possible, they usually result in overly complex expressions, highlighting the need for automated approaches.

Simplification techniques are essential for making symbolic expressions interpretable, allowing designers to identify key components and understand circuit behavior more deeply [9]. Based on the timing of simplification, methods are commonly categorized into SBG, SDG, and SAG [10]. SBG methods simplify circuit matrices, graphs, or models before symbolic generation, often using matrix-, graph-, or circuit-based strategies [11,12]. While SBG can reduce the complexity of subsequent symbolic results, some approaches may yield slightly more complex expressions to preserve correspondence with the original circuit. SDG techniques generate simplified symbolic expressions directly during analysis, prioritizing terms by influence on circuit behavior and using error control mechanisms such as coefficient-, sensitivity-, or sampling-based checks [13]. These methods offer faster computation and lower memory use, making them suitable for larger circuits. In contrast, SAG techniques simplify fully derived symbolic network functions by pruning insignificant terms from expanded polynomials [14–17]. While effective, SAG approaches can have limited accuracy outside nominal points, and more advanced criteria have been proposed to manage pole/zero displacements and control cumulative errors.

Recently, metaheuristic algorithms have gained attention for symbolic simplification [6,18–21]. Evolutionary methods such as GA, simulated annealing (SA), and hybrid approaches have been applied to generate simplified expressions for PSRR, noise analysis, and other performance metrics while preserving accuracy [18,19,21]. Grey wolf optimizer (GWO) has also been used to optimize analog circuits such as differential amplifiers and operational amplifiers [20]. Similarly, a hybrid metaheuristic algorithm based on artificial bee colony and SA (named ABC-SA) have been employed for simplified pole/zero extraction and transfer function factorization [6]. These methods aim to identify the most critical circuit components and reduce symbolic complexity, offering more manageable expressions for designers.

Building on these advances, this paper introduces the DSSA tool, which departs from conventional graph- or matrix-based methods. DSSA treats symbolic simplification as a modeling problem, directly producing only the most significant terms of the transfer function without requiring prior full symbolic analysis. This approach efficiently reduces computational complexity and memory requirements, providing a practical solution for real-size, transistor-level circuits and extending the applicability of symbolic analysis in modern analog design.

## 3. Proposed DSSA Method

Let us consider the symbolic circuit transfer function in an expanded form as a function of the complex frequency $s$ and the vector of $K$ circuit parameters $\mathbf{x} = (x_1, x_2, \ldots, x_K)$ as follows:

$$H(s, \mathbf{x}) = \frac{f_0(\mathbf{x}) + sf_1(\mathbf{x}) + s^2 f_2(\mathbf{x}) + \cdots + s^M f_M(\mathbf{x})}{g_0(\mathbf{x}) + sg_1(\mathbf{x}) + s^2 g_2(\mathbf{x}) + \cdots + s^N g_N(\mathbf{x})}, \tag{1}$$

where the polynomials $f_i(\mathbf{x})$ or $g_j(\mathbf{x})$ is a sums-of-products of $\mathbf{x}$, which can be calculated as follows:

$$h(\mathbf{x}) = \sum_{t=1}^{T} h_t(\mathbf{x}) = h_1(\mathbf{x}) + h_2(\mathbf{x}) + \ldots + h_T(\mathbf{x}). \tag{2}$$

Since each symbolic term $h_t$ is a product of $\mathbf{x}$, the simplification problem can be viewed as a modeling problem to directly generate the most significant terms, so that approximate polynomials are achieved. It is assumed that each symbolic circuit parameter (circuit element or device model parameter) can take any value within a variation range, i.e., $x_k \in [L_k, H_k]$, where, $L_k$ and $H_k$ represent the lower and upper bounds for the circuit parameter $x_k$, respectively.

Based on the range of circuit parameters, a dataset comprising $D$ data points is collected, where the value of the $k$-th circuit parameter within the $d$-th data point ($d$=1, 2, …, $D$) is randomly achieved as $x_k^d = L_k + r(H_k - L_k)$, and $r \in [0,1]$ is a random number with uniform distribution. The full dataset is randomly separated into train and test datasets with $D_{train}$ and $D_{test}$ data points, respectively.

### 3.1. Solution encoding/decoding

In DSSA, the circuit is numerically analysed via the modified nodal analysis, resulting in the exact numeric transfer function. Then, GA is applied to generate the simplified transfer function. As seen in Figure 1, each feasible solution (i.e., each chromosome within the GA) is encoded via a matrix of dimension $P \times Q$, where $P$ ($P$=$M$+$N$+2) is the number of all polynomials within both the numerator and denominator of the symbolic transfer function, and $Q$ is the number of genes within each polynomial of the transfer function. Here, $Q$=$T \times (K+1)$, where $K$ is the number of all symbolic parameters of the circuit, and $T$ is the user-specified maximum number of terms for each polynomial.

To decode each feasible solution, each symbolic term can be constructed as a string of length ($K$+1). For instance, in Figure 1, the highlighted genes illustrate the initial term of polynomial $f_0$, consisting of $K$+1 genes. The first $K$ genes denote the presence or absence of various parameters within the symbolic term, while the final gene serves as the term selector, indicating the inclusion or exclusion of the overall symbolic term as well as its sign in the corresponding polynomial. More specifically, the $K$ first genes have binary values; $S(p,(t-1) \times (K+1)+k)$=1, if the $k$-th parameter presents in the $t$-th term of the $p$-th polynomial, and it absents for the otherwise. Furthermore, the last gene is the term selector, $TS(p,t)$, defining the presence or absence of the symbolic term $t$ within the corresponding polynomial $p$. The value of $TS(p,t)$ can be +1, -1, or 0, for presence with a positive sign, presence with a negative sign, or absence of the $t$-th term in the $p$-th polynomial, respectively.

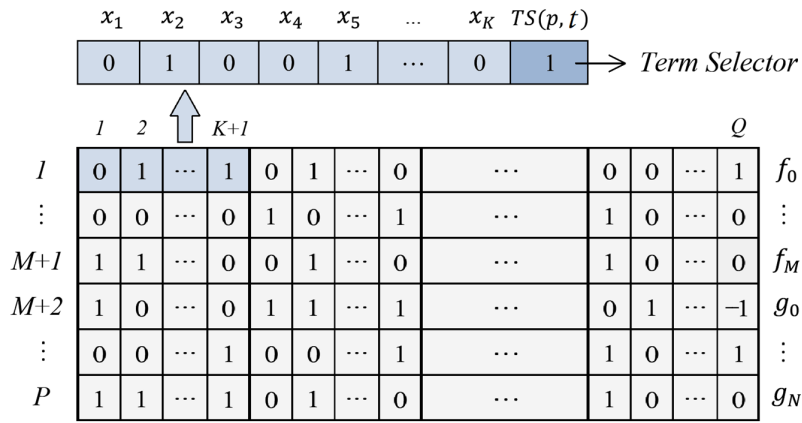

**Figure 1.** Solution representation (a feasible chromosome encoding).

## 3.2. Objective function

To evaluate a solution, the numerical results of the simplified solution (corresponding to each chromosome) are compared with the exact ones at every train data point. The aim of the objective function is to minimize the total expression complexity (*Complexity*) and magnitude/phase error (*Error*). These multiple objectives are merged into a single-objective function as follows:

$$\text{minimize } Obj = w_1 \times Complexity + w_2 \times Error \tag{3}$$

subject to

$$|\Delta H(0, \mathbf{x}_d)| \leq T_{dc} \quad \forall d, \tag{4}$$

$$\left|\left(p_{n,d}^{new} - p_{n,d}^{exact}\right)/p_{n,d}^{exact}\right| \leq T_{root} \quad \forall n, \forall d, \tag{5}$$

$$\left|\left(z_{m,d}^{new} - z_{m,d}^{exact}\right)/z_{m,d}^{exact}\right| \leq T_{root} \quad \forall m, \forall d, \tag{6}$$

where $T_{dc}$ and $T_{root}$ are maximum-allowable errors for dc-gain error and pole/zero displacements, respectively. $\Delta H(0, \mathbf{x}_d)$ is dc-gain error, $p_{n,d}$ is $n$-th pole ($n=1,2,\ldots,N$), $z_{m,d}$ is $m$-th zero ($m=1,2,\ldots,M$), for $d$-th train data ($d=1,2,\ldots,D_{train}$), and $T_{dc}$ and $T_{root}$ are set according to the required accuracy level.

$$Complexity = \frac{1}{P \times T} \sum_{p=1}^{P} \sum_{t=1}^{T} |TS(p,t)|, \tag{7}$$

$$Error = \frac{1}{2D_{train}C} \sum_{d=1}^{D_{train}} \sum_{c=1}^{C} \left(|\Delta H(s_c, \mathbf{x}_d)| + |\Delta \varphi(s_c, \mathbf{x}_d)|\right), \tag{8}$$

where $P$ represents the collective number of all polynomials contained in both the numerator and denominator of the transfer function, and $T$ signifies the maximum count of symbolic terms within each polynomial. Additionally, the notation $|TS(p,t)|$ is assigned a value of 1 if the $t$-th term is present in the $p$-th polynomial; conversely, it is set to 0 if the term is absent. Furthermore, $\Delta H(s_c, \mathbf{x}_d)$ and $\Delta \varphi(s_c, \mathbf{x}_d)$ are the error in magnitude and phase in the $c$-th frequency point for the $d$-th train data point, respectively.

## 3.3. Optimization procedure

Search methods are generally classified as exact, heuristic, or metaheuristic approaches [22]. Exact techniques guarantee optimal solutions but are computationally impractical for large circuits. Heuristics, while fast and simple, explore only part of the search space, whereas metaheuristics strike a balance, offering high-quality solutions within reasonable time [23]. The choice of metaheuristic depends on the problem type: some excel in combinatorial optimization, others in continuous domains [24]. Among combinatorial problems, the genetic algorithm (GA) is the most widely used, as reflected in numerous studies [25–33]. Accordingly, GA was selected here to optimize the DSSA model.

The GA optimization begins with a randomly generated population. In each iteration, the population is evaluated using an objective function, followed by updates through crossover and mutation. In crossover, two parents are selected via roulette wheel selection, and their genetic information is combined using uniform crossover. During mutation, a parent is chosen using RSW selection, and a gene is randomly swapped to produce offspring.

## 4. Simulation results

### 4.1. Settings

The proposed DSSA method has been developed in MATLAB. The program was exploited to derive the simplified transfer function of a three-stage nested-Miller amplifier in the $RCg_m$ model (NMAM) as seen in Figure 2, and four CMOS amplifiers including a two-stage Miller amplifier (MA) [18], a three-stage nested-Miller amplifier (NMA) [19], a hybrid-cascade amplifier (HCA) [34] and a folded-cascade amplifier (FCA) [35]. Details of the study circuits can be found in [1]. The number of nodes, parameters and exact terms within each circuit are summarized in Table 1.

For setting the GA parameters, population size, the maximum number of iterations, recombination probability ($P_R$), crossover probability ($P_C$), and mutation probability ($P_M$), were set to 50, 1000, 0.1, 0.5, and 0.4, respectively. In our simulations, 3 db and 30% were considered for $T_{dc}$ and $T_{root}$, respectively. We set the maximum terms within each polynomial as $T$=15. The weights of the objective function were set as $w_1$=0.8 and $w_2$=0.2. Three frequency points are used at each frequency decade. In each circuit, 100 train and 50 test data points were generated. Each circuit parameter is randomly determined within ±50% variation around its nominal point.

**Table 1.** Details of the study circuits.

| Circuit | No. Nodes | No. Parameters | No. Exact Terms |
|---|---|---|---|
| *NMAM* | 4 | 11 | 40 |
| *MA* | 9 | 26 | 134 |
| *HCA* | 14 | 27 | 6804 |
| *NMA* | 13 | 36 | 9820 |
| *FCA* | 14 | 45 | 11484 |

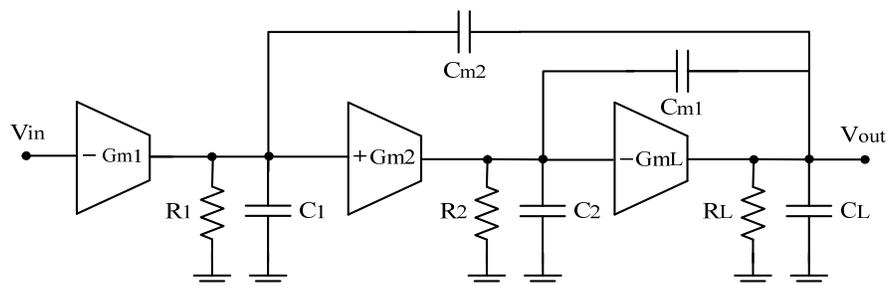

**Figure 2.** A typical amplifier with nested-Miller compensation (NMAM).

### 4.2. Results

In this section, the performance of the proposed DSSA method is evaluated against three single-point simplification techniques: a topological method (SP1) [36], a GA-based SAG approach (SP2) [19], a factorization method (SP3) [6], and a variation range-based metaheuristic method (VR) [37]. Table 2 compares these approaches in terms of the total number of symbolic terms generated during analysis and the number of final simplified terms. Tables 3 and 4 report the average and maximum errors in dc-gain and pole/zero locations across the test datasets, while Figures 3 and 4 summarize these results averaged over all circuits. The results clearly highlight the effectiveness of DSSA in minimizing errors across all test points. Compared to the other techniques,

DSSA consistently produces more accurate representations of dc-gain and pole/zero positions, reflecting its ability to capture the most significant terms of the transfer function.

A key advantage of DSSA is its efficiency: it generates far fewer symbolic terms during analysis, particularly for larger circuits, which substantially reduces memory usage and computational overhead. This streamlined term generation also enables the derivation of more compact expressions with lower complexity, simplifying further analysis and interpretation.

The results demonstrate that DSSA not only improves accuracy but also enhances practicality for real-world circuit analysis, providing a reliable and memory-efficient alternative to conventional simplification techniques. Its balance of precision and efficiency makes it particularly valuable for engineers and researchers dealing with complex analog circuits.

**Table 2.** Comparison of the number of terms obtained by different techniques.

| Circuit | Number of all generated terms | | | | | Number of simplified terms | | | | |
|---|---|---|---|---|---|---|---|---|---|---|
| | *SP1* | *SP2* | *SP3* | *VR* | *DSSA* | *SP1* | *SP2* | *SP3* | *VR* | *DSSA* |
| *NMAM* | 40 | 80 | 40 | 40 | 115 | 20 | 15 | 8 | 17 | 12 |
| *MA* | 134 | 268 | 134 | 134 | 75 | 26 | 19 | 8 | 25 | 28 |
| *HCA* | 6804 | 15348 | 6804 | 6804 | 115 | 44 | 35 | 14 | 82 | 30 |
| *NMA* | 9820 | 19640 | 9820 | 9820 | 115 | 95 | 76 | 18 | 127 | 42 |
| *FCA* | 11484 | 28339 | 11484 | 11484 | 75 | 92 | 50 | 9 | 119 | 38 |

**Table 3.** Comparison of the dc-gain error (db) obtained by different techniques.

| Circuit | Average dc-gain error (db) | | | | | Maximum dc-gain error (db) | | | | |
|---|---|---|---|---|---|---|---|---|---|---|
| | *SP1* | *SP2* | *SP3* | *VR* | *DSSA* | *SP1* | *SP2* | *SP3* | *VR* | *DSSA* |
| *NMAM* | 0 | 0 | 0 | 0 | 0 | 0 | 0 | 0 | 0 | 0 |
| *MA* | 0.6 | 1.7 | 1.1 | 0.1 | 0.1 | 2.2 | 3.1 | 3.6 | 0.2 | 0.2 |
| *HCA* | 0.9 | 1.6 | 2.7 | 0.9 | 0.8 | 2.5 | 3.2 | 7.5 | 2.1 | 2.4 |
| *NMA* | 1.8 | 2.1 | 2.2 | 1.1 | 1.2 | 3.9 | 4.7 | 8.8 | 1.7 | 2.5 |
| *FCA* | 2.5 | 3.2 | 6.9 | 1.5 | 1.1 | 6.3 | 8.1 | 12.5 | 2.9 | 1.7 |

**Table 4.** Comparison of the pole/zero error (%) obtained by different techniques.

| Circuit | Average pole/zero error (%) | | | | | Maximum pole/zero error (%) | | | | |
|---|---|---|---|---|---|---|---|---|---|---|
| | *SP1* | *SP2* | *SP3* | *VR* | *DSSA* | *SP1* | *SP2* | *SP3* | *VR* | *DSSA* |
| *NMAM* | 1 | 10 | 13 | 1 | 1 | 4 | 58 | 37 | 4 | 5 |
| *MA* | 17 | 22 | 16 | 11 | 9 | 28 | 55 | 48 | 25 | 18 |
| *HCA* | 11 | 17 | 8 | 8 | 10 | 32 | 40 | 34 | 21 | 23 |
| *NMA* | 7 | 26 | 15 | 6 | 8 | 17 | 47 | 42 | 16 | 21 |
| *FCA* | 11 | 29 | 19 | 10 | 6 | 25 | 53 | 45 | 21 | 17 |

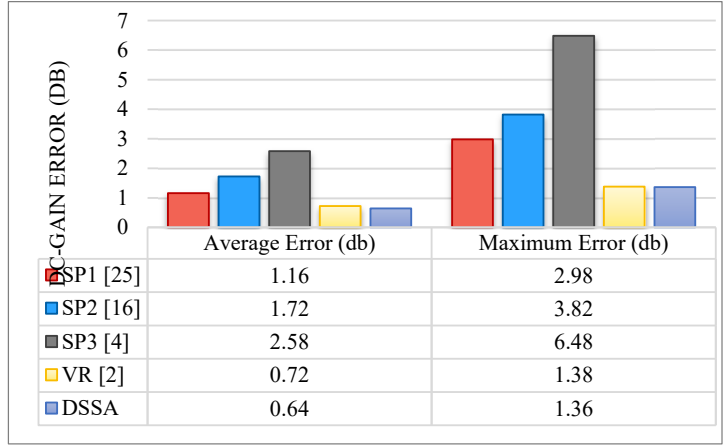

**Figure 3.** Comparison of the average and maximum dc-gain error.

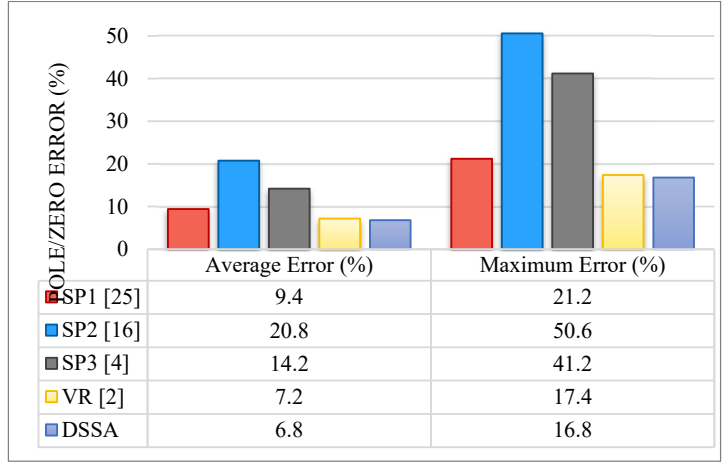

**Figure 4.** Comparison of the average and maximum pole/zero error.

## 5. Conclusion

This paper presents Direct Simplified Symbolic Analysis (DSSA), a fast and accurate method for the symbolic analysis of analog circuits. DSSA significantly reduces the number of generated symbolic terms, lowering memory requirements and enabling efficient analysis of large circuits. By accounting for parameter variations, its simplified expressions are more generally valid than single-point approaches, making it valuable for engineers and researchers working on complex analog designs. The method also opens promising avenues for circuit optimization and automated synthesis.

In this paper, Monte Carlo simulation has been used to handle parameter uncertainties and generate datasets for model training and testing. Future work could explore robust optimization [38], stochastic programming [22], Mamdani [39] and Takagi-Sugeno fuzzy logic [40], to better address parameter variations. Additionally, other metaheuristics, including ant colony optimization [41], artificial bee colony [42–44], or ensemble heuristic-metaheuristic algorithms [45], could enhance the generation of simplified expressions. Further directions may involve inverse problem theory [46] or machine learning and deep learning methods [47,48] to capture circuit behavioral patterns.

# References

[1] Shokouhifar, M., Yazdanjouei, H., & Weber, G. W. (2024). DSSA: Direct Simplified Symbolic analysis using metaheuristic-driven circuit modelling. *Journal of Dynamics and Games*, *11*(3), 232-248.

[2] Shokouhifar, M., & Jalali, A. (2014, May). Automatic symbolic simplification of analog circuits in MATLAB using ant colony optimization. In *2014 22nd Iranian Conference on Electrical Engineering (ICEE)* (pp. 407-412). IEEE.

[3] Fang, W., & Ying, M. (2024, June). Symphase: Phase symbolization for fast simulation of stabilizer circuits. In *Proceedings of the 61st ACM/IEEE Design Automation Conference* (pp. 1-6).

[4] Shi, G., Tan, S. X. D., & Tlelo-Cuautle, E. (2014). Advanced symbolic analysis for VLSI systems. *Methods and Applications, Berlin, Springer*.

[5] Bahadoran, G., & Reza-Alikhani, H. (2021). Four-stage CMOS amplifier design based on symbolic calculations: Stack comparison approach. *International Journal of Numerical Modelling: Electronic Networks, Devices and Fields*, *34*(4), e2867.

[6] Shokouhifar, M., & Jalali, A. (2017). Simplified symbolic transfer function factorization using combined artificial bee colony and simulated annealing. *Applied Soft Computing*, *55*, 436-451.

[7] Zhao, Z., & Zhang, L. (2020). An automated topology synthesis framework for analog integrated circuits. *IEEE Transactions on Computer-Aided Design of Integrated Circuits and Systems*, *39*(12), 4325-4337.

[8] Aminzadeh, H., Grasso, A. D., & Palumbo, G. (2022). A methodology to derive a symbolic transfer function for multistage amplifiers. *IEEE Access*, *10*, 14062-14075.

[9] Filaretov, V., & Gorshkov, K. (2020). Efficient generation of compact symbolic network functions in a nested rational form. *International Journal of Circuit Theory and Applications*, *48*(7), 1032-1056.

[10] Tan, S. D., & Shi, C. J. (2004). Efficient approximation of symbolic expressions for analog behavioral modeling and analysis. *IEEE Transactions on Computer-Aided Design of Integrated Circuits and Systems*, *23*(6), 907-918.

[11] Fernández, F. V., López, C. S., Castro-López, R., & Roca-Moreno, E. (2012). Approximation Techniques in Symbolic Circuit Analysis. *Design of Analog Circuits through Symbolic Analysis*, 173-201.

[12] Hayes, M. (2022). Lcapy: Symbolic linear circuit analysis with Python. *PeerJ Computer Science*, *8*, e875.

[13] Lai, X., Wang, S., Ma, S., Xie, J., & Zheng, Y. (2020). Parameter sensitivity analysis and simplification of equivalent circuit model for the state of charge of lithium-ion batteries. *Electrochimica Acta*, *330*, 135239.

[14] Wierzba, G. M., Srivastava, A., Joshi, V., Noren, K. V., & Svoboda, J. A. (1989, August). SSPICE-A symbolic SPICE program for linear active circuits. In *Proceedings of the 32nd Midwest Symposium on Circuits and Systems,* (pp. 1197-1201). IEEE.

[15] Rutenbar, R. A., Gielen, G. G., & Antao, B. A. (2002). Interactive AC Modeling and Characterization of Analog Circuits via Symbolic Analysis.

[16] Gielen, G. G., Walscharts, H. C., & Sansen, W. M. (2002). ISAAC: A symbolic simulator for analog integrated circuits. *IEEE Journal of solid-state circuits*, *24*(6), 1587-1597.

[17] Fakhfakh, M., Tlelo-Cuautle, E., & Fernández, F. V. (Eds.). (2012). *Design of analog circuits through symbolic analysis*. Bentham Science Publishers.

[18] Shokouhifar, M., & Jalali, A. (2016). Evolutionary based simplified symbolic PSRR analysis of analog integrated circuits. *Analog Integrated Circuits and Signal Processing*, *86*(2), 189-205.

[19] Shokouhifar, M., & Jalali, A. (2015). An evolutionary-based methodology for symbolic simplification of analog circuits using genetic algorithm and simulated annealing. *Expert Systems with Applications*, *42*(3), 1189-1201.

[20] Majeed, M. M., & Rao, P. S. (2017, December). Optimization of CMOS analog circuits using grey wolf optimization algorithm. In *2017 14th IEEE India Council International Conference (INDICON)* (pp. 1-6). IEEE.

[21] Panda, M., Patnaik, S. K., & Mal, A. K. (2021). An efficient method to compute simplified noise parameters of analog amplifiers using symbolic and evolutionary approach. *International Journal of Numerical Modelling: Electronic Networks, Devices and Fields*, *34*(1), e2790.

[22] Shokouhifar, M., Hasanvand, M., Moharamkhani, E., & Werner, F. (2024). Ensemble heuristic–metaheuristic feature fusion learning for heart disease diagnosis using tabular data. *Algorithms*, *17*(1), 34.

[23] Shokouhifar, M., & Pilevari, N. (2022). Combined adaptive neuro-fuzzy inference system and genetic algorithm for E-learning resilience assessment during COVID-19 Pandemic. *Concurrency and Computation: Practice and Experience*, *34*(10), e6791.

[24] Sabet, S., Shokouhifar, M., & Farokhi, F. (2016). A comparison between swarm intelligence algorithms for routing problems. *Electrical & Computer Engineering: An International Journal (ECIJ)*, *5*(1), 17-33.

[25] Yang, J., Shokouhifar, M., Yee, L., Khan, A. A., Awais, M., & Mousavi, Z. (2024). DT2F-TLNet: A novel text-independent writer identification and verification model using a combination of deep type-2 fuzzy architecture and Transfer Learning networks based on handwriting data. *Expert Systems with Applications*, *242*, 122704.

[26] Shokouhifar, M., & Hassanzadeh, A. (2014). An energy efficient routing protocol in wireless sensor networks using genetic algorithm. *Advances in Environmental Biology*, *8*(21), 86-93.

[27] Nahavandi, B., Homayounfar, M., Daneshvar, A., & Shokouhifar, M. (2022). Hierarchical structure modelling in uncertain emergency location-routing problem using combined genetic algorithm and simulated annealing. *International Journal of Computer Applications in Technology*, *68*(2), 150-163.

[28] Sohrabi, M., Fathollahi-Fard, A. M., & Gromov, V. A. (2024). Genetic engineering algorithm (GEA): an efficient metaheuristic algorithm for solving combinatorial optimization problems. *Automation and Remote Control*, *85*(3), 252-262.

[29] Ruan, Y., Cai, W., & Wang, J. (2024). Combining reinforcement learning algorithm and genetic algorithm to solve the traveling salesman problem. *The Journal of Engineering*, *2024*(6), e12393.

[30] Danach, K., Kanj, H., Hamze, K., & Moukadem, I. (2024). Optimizing Learning-Based Combinatorial Optimization Algorithms: Advanced Hyperparameter Techniques and Real-World Applications. *Nanotechnol Percept*, *20*(S15), 2996-3017.

[31] Tang, Z., Wang, L., Guo, S., Liang, G., Zhang, W., Zhang, L., ... & Wang, Y. (2025). Study on modular design methodology of marine SMR system based on fuzzy hierarchical clustering and improved genetic algorithm. *Progress in Nuclear Energy*, *185*, 105739.

[32] Ye, M. (2025, February). Application of Genetic Algorithms in Combinatorial Optimization Problems and Algorithm Analysis. In *2025 3rd International Conference on Integrated Circuits and Communication Systems (ICICACS)* (pp. 1-5). IEEE.

[33] Gu, S., Li, K., Xing, J., Zhang, Y., & Cheng, J. (2025). Synergizing Reinforcement Learning and Genetic Algorithms for Neural Combinatorial Optimization. *arXiv preprint arXiv:2506.09404*.

[34] Yavari, M. (2005). Hybrid cascode compensation for two-stage CMOS opamps. *IEICE transactions on electronics*, *88*(6), 1161-1165.

[35] Akbari, M., & Hashemipour, O. (2014). Enhancing transconductance of ultra-low-power two-stage folded cascode OTA. *Electronics Letters*, *50*(21), 1514-1516.

[36] Hu, H., Shi, G., Tai, A., & Lee, F. (2015, May). Topological symbolic simplification for analog design. In *2015 IEEE International Symposium on Circuits and Systems (ISCAS)* (pp. 2644-2647). IEEE.

[37] Shokouhifar, M., & Jalali, A. (2017). Variation range based simplification for meaningful symbolic analysis of analog integrated circuits. *Analog Integrated Circuits and Signal Processing*, *90*(2), 447-461.

[38] Özmen, A., Kropat, E., & Weber, G. W. (2017). Robust optimization in spline regression models for multi-model regulatory networks under polyhedral uncertainty. *Optimization*, *66*(12), 2135-2155.

[39] Carrasco-Garrido, C., Moreno-Cabezali, B. M., & Martínez Raya, A. (2025). New perspectives on university quality assessment: A Mamdani Fuzzy Inference System approach. *PloS one*, *20*(5), e0321013.

[40] Memarian, S., Behmanesh-Fard, N., Aryai, P., Shokouhifar, M., Mirjalili, S., & del Carmen Romero-Ternero, M. (2024). TSFIS-GWO: Metaheuristic-driven takagi-sugeno fuzzy system for adaptive real-time routing in WBANs. *Applied Soft Computing*, *155*, 111427.

[41] Shokouhifar, M., & Sabet, S. (2012, July). PMACO: A pheromone-mutation based ant colony optimization for traveling salesman problem. In *2012 International Symposium on Innovations in Intelligent Systems and Applications* (pp. 1-5). IEEE.

[42] Shokouhifar, M., & Farokhi, F. (2010, December). An artificial bee colony optimization for feature subset selection using supervised fuzzy C_means algorithm. In *3rd International conference on information security and artificial intelligent (ISAI)* (pp. 427-432).

[43] Sabet, S., Shokouhifar, M., & Farokhi, F. (2013). A discrete artificial bee colony for multiple knapsack problem. *International Journal of Reasoning-based Intelligent Systems*, *5*(2), 88-95.

[44] Shokouhifar, M., & Sabet, S. (2010, December). A hybrid approach for effective feature selection using neural networks and artificial bee colony optimization. In *3rd international conference on machine vision (ICMV 2010)* (pp. 502-506).

[45] Behmanesh-Fard, N., Yazdanjouei, H., Shokouhifar, M., & Werner, F. (2023). Mathematical Circuit Root Simplification Using an Ensemble Heuristic–Metaheuristic Algorithm. *Mathematics*, *11*(6), 1498.

[46] Aster, R. C., Borchers, B., & Thurber, C. H. (2018). *Parameter estimation and inverse problems*. Elsevier.

[47] Shokouhifar, A., Shokouhifar, M., Sabbaghian, M., & Soltanian-Zadeh, H. (2023). Swarm intelligence empowered three-stage ensemble deep learning for arm volume measurement in patients with lymphedema. *Biomedical Signal Processing and Control*, *85*, 105027.

[48] Yang, J., Shokouhifar, M., Yee, L., Khan, A. A., Awais, M., & Mousavi, Z. (2024). DT2F-TLNet: A novel text-independent writer identification and verification model using a combination of deep type-2 fuzzy architecture and Transfer Learning networks based on handwriting data. *Expert Systems with Applications*, *242*, 122704.